\documentclass[12pt]{article}

\usepackage{amsfonts,epsfig,amsmath,amsthm,amssymb,graphics,verbatim,overpic}
\usepackage[letterpaper,margin=0.85in]{geometry}

\def\R{\mathbb{R}}

\def\N{\mathbb{N}}

\def\I{\infty}

\newcommand{\be}{\begin{equation}}
\newcommand{\ee}{\end{equation}}
\newcommand{\bea}{\begin{eqnarray}}
\newcommand{\eea}{\end{eqnarray}}
\newcommand{\beann}{\begin{eqnarray*}}
\newcommand{\eeann}{\end{eqnarray*}}
\newcommand{\benn}{\begin{equation*}}
\newcommand{\eenn}{\end{equation*}}

\def\ra{\rightarrow}
\def\I{\infty}

\newcommand{\cD}{{\mathcal D}}  
\newcommand{\cF}{{\mathcal F}}  
\newcommand{\cL}{{\mathcal L}}  
\newcommand{\cX}{{\mathcal X}}  

\def\txtd{{\textnormal{d}}}
\def\txte{{\textnormal{e}}}
\def\txti{{\textnormal{i}}}

\def\txtD{{\textnormal{D}}}

\begin{document}

\author{Christian Kuehn\thanks{Technical University of Munich, Faculty of Mathematics,
Boltzmannstr.~3, 85748 Garching b.~M{\"u}nchen, Germany; e-mail: ckuehn@ma.tum.de}}
 
\title{Network Dynamics on Graphops}

\maketitle

\begin{abstract}
In this brief note, we report a formal mathematical observation: we are about to 
breach a major century-old barrier in the analysis of interacting particle systems.
More precisely, it is well-known that in well-mixed/homogeneous/all-to-all-coupled systems, 
one may derive mean-field limit equations such as Vlasov-Fokker-Planck equations (VFPEs). A
mesoscopic VFPE describes the probability of finding a single vertex/particle in a 
certain state, forming a bridge between microscopic statistical physics and macroscopic
fluid-type approximations. One major obstacle in this framework is to incorporate complex
network structures into limiting equations. In many cases, only heuristic approximations 
exist, or the limits rely on particular classes of integral operators. In this paper,
we notice that there is a much more elegant, and profoundly more general, way available
due to recent progress in the theory of graph limits. In particular, we show how one
may easily enter complex network dynamics via graphops (graph operators) into VFPEs.   
\end{abstract}


\section{Introduction}
\label{sec:intro}

Interacting particle systems, or more generally, dynamical systems on graphs/networks, form 
one of the major building blocks of modern science~\cite{WattsStrogatz,BarabasiAlbert}. Within 
the last three decades, they have permeated virtually all disciplines, ranging from 
molecular scales~\cite{Feinberg} to neuroscience~\cite{Sporns}, systems biology~\cite{Palsson}, 
machine learning~\cite{Haykin}, social science~\cite{Scott2}, 
epidemiology~\cite{BarratBarthelemyVespignani}, and transportation 
networks~\cite{RanBoyce} up to climate science scales~\cite{Dongesetal}. For all-to-all coupled 
systems, lattice systems, and various special classes of network dynamics with similar dynamics 
at each vertex, there is a well-developed theory to pass to a mean-field 
approximation in many theoretical frameworks~\cite{Golse,KissMillerSimon,Kuramoto,LasryLions,
Meleard,Spohn1}. One considers the limit of an infinite graph~\cite{vanderHofstad} 
\benn
n\ra \I
\eenn
so the particle/vertex number tends to infinity. If one is interested in the probability 
density $\rho=\rho(u,t)$ to find a vertex in a certain dynamical state $u\in\R^k$ at time 
$t$, one may often derive a differential equation for $\rho$. One common form for $k=1$
is
\be
\label{eq:VFPEintro}
\partial_t = - \partial_u (\rho V(\rho)),
\ee 
where the map $V$ can be derived from the dynamics of each vertex~\cite{Golse,Frank}. If each 
vertex also is influenced by Gaussian noise, a second-order term $\partial_{uu}(\cdot)$ 
commonly appears as well in~\eqref{eq:VFPEintro}. The equation~\eqref{eq:VFPEintro} is sometimes 
referred to as Liouville equation, continuity equation~\cite{Strogatz1}, and/or Vlasov equation, 
while the second-order equation is more known as Fokker-Planck and/or Kolmogorov equation. Here 
we shall refer to this class of equations as Vlasov-Fokker-Planck equation (VFPEs).\medskip 

Now it is evident that one would like to derive a far more general form 
of~\eqref{eq:VFPEintro}, which also takes into account cases, where the adjacency
matrix $A=A^{(n)}\in\R^{n\times n}$ of the interaction between finitely many vertices 
is not just the the matrix of a full graph nor a highly symmetric structure such as a lattice. 
There have been many recent steps forward to achieve this goal. Several mathematical 
approaches have been successful providing rigorous proofs for VFPEs, where nonlocal 
integral terms appear to take into account the heterogeneous coupling 
structure~\cite{ChibaMedvedev,LasryLions}. 
However, a general theory, which would allow us to upgrade easily from particular cases or 
standard all-to-all mean-field limit VFPEs, to modern complex network structures is still 
lacking.\medskip 

In this note, we propose that a path to achieve an elegant extension of
all classical approaches is to incorporate an analytical approach to graph limit theory.
Graph limit theory~\cite{Lovasz}, i.e., taking $n\ra \I$, has made
significant progress in recent years. New limit structures for dense graphs via 
graphons~\cite{LovaszSzegedy}, as well as for sparse 
graphs~\cite{BollobasRiordan,Borgsetal,HatamiLovaszSzegedy}, have recently 
appeared. The theory is relatively technical and convergence notions are often
hard to understand as they have relied on combinatorial structures~\cite{Lovasz}. Recently a
new approach to unify and extend graph limit theory was proposed by Backhausz and
Szegedy~\cite{BackhauszSzegedy}. Their idea relies on viewing the adjacency matrix
$A^{(n)}$ from an operator-theoretic perspective. In the simplest setting, 
one takes a vector $v$ in $\R^n$ and consider the $2\times n$ matrix formed from
$v$ and $v^\top A^{(n)}$. Sampling the columns of this matrix uniformly at random
generates a measure $\mu_v$. Hence, one may view graphs also via their associated
measures. Different graphs, even of different sizes, can then be more easily 
compared as we only view them through their \emph{action}; see Section~\ref{sec:graphops}
for more details and a brief review of the class of graphops (graph operators), which
we shall rely on. Backhausz and Szegedy show that in a suitably chosen metric, large
classes of graphs do have non-trivial subsequential limits
\benn
A^{(n)}\ra A^{(\I)}\qquad \text{as $n\ra \I$},
\eenn 
where $A^{(\I)}$ can also be viewed as an operator, but now acting as an abstract 
operator on a function space. We are going to show (formally) that exactly this 
viewpoint is the missing ingredient to start a general theory of VFPEs, and related 
classes of evolution equations, as limits of finite-dimensional network dynamics.\medskip 

We are going to illustrate our reasoning primarily on a benchmark class of 
interacting particle systems, the famous Kuramoto model of coupled 
oscillators~\cite{Kuramoto,Strogatz1}. This model is widely studied and has served as 
a benchmark case to understand self-organizing and adaptive nonlinear synchronization. 
It can be embedded into a more general class of kinetic-type models. We shall see, how these 
kinetic models also fit elegantly together with graphops. This approach really seems to 
break a barrier that has held back the application of tools from dynamical 
systems, functional analysis, and evolution equations to broad classes of 
large-scale network dynamics problems.\medskip

The paper is structured as follows: In Section~\ref{sec:graphops}, we review the
relevant parts of graphop theory. In Section~\ref{sec:micro}, we show how this
theory also provides an elegant framework to deal with network dynamics on a 
finite-dimensional level, i.e., for finite graphs. For example, we show that
the graphop viewpoint provides a completely natural way to intertwine in one set
of equations the dynamics on the network and the correlation structure between
network elements. Then we proceed to formally develop, why graphops do generalize
previous approaches to VFPEs in Section~\ref{sec:meso}. We conclude in 
Section~\ref{sec:outlook} with an outlook towards future opportunities and
challenges. 
      
\section{Graphs as Operators}
\label{sec:graphops}

We build upon the theory of graphops from~\cite{BackhauszSzegedy}. Let $(\Omega,\cF,\mu)$
be a probability space and let $L^p(\Omega,\R)=L^p(\Omega)$ denote the usual Lebesgue 
space for $p\in [1,\I]$. A linear operator $A:L^\I(\Omega)\ra L^1(\Omega)$ is called 
\emph{P-operator} if the operator norm
\benn
\|A\|_{\infty \ra 1}:=\sup_{v\in L^\I(\Omega)}\frac{\|Av\|_1}{\|v\|_\I}
\eenn
is finite. Key examples of P-operators are matrices, which we want to view as adjacency 
matrices of graphs. Indeed, consider $\Omega=\{1,2,\ldots,n\}=:[n]$
for $n\in\N$ with the uniform measure $\mu_{[n]}$ and sigma-algebra given by the 
power set $2^{[n]}$. Then a vector $v\in\R^n$ defines a function $v:[n]\ra \R$ 
by $v(j)=v_j$ for $j\in[n]$. On $([n],2^{[n]},\mu_{[n]})$ any matrix $A\in\R^{n\times n}$ 
is a P-operator acting on (row) vectors via the usual rule 
\benn
(vA)(j)=\sum_{k=1}^n v(k)A_{kj}=\sum_{k=1}^n v_kA_{kj}.
\eenn
Obviously, we can identify $L^\I([n])$ and $L^1([n])$ with $\R^n$ and finite-dimensional 
linear operators given by matrices are bounded in the associated operator norm. 

A key advantage of general P-operators are their convergence properties, when one
considers sequences of these operators via \emph{profiles}. As an example, we 
consider again $A\in\R^{n\times n}$ and any (row) vector $v\in\R^n$, then we have $vA
\in \R^{1\times n}$ so that we may form a matrix $M\in \R^{2\times n}$ with rows $v$ 
and $vA$. Now we sample columns of $M$ uniformly, which yields a probability measure 
$\mu_v$ on $\R^2$. The (1-)profile of $A$ is given by the collection of measures 
$\{\mu_v:v\in\R^2\}$. One may generalize the notion of a $1$-profile to a $k$-profile 
$\mathcal{S}_k(A)$ for a general P-operator where one uses $k$ vectors $v_1,v_2,\ldots,v_k$ 
as the first $k$ rows and $v_1 A,v_2 A,\ldots,v_k A$ as $k$ further rows. Hence, the 
matrix $M$ becomes a matrix of size $\R^{2k\times n}$; wlog one may restrict 
$v_j\in [-1,1]^n$ as the results do not change upon scaling vectors. A random column 
in $M$ yields a probability measure on $\R^{2k}$ and $\mathcal{S}_k(A)$ is the collection 
of all such measures. 

For a general P-operator $A$, consider functions $v_{1},\ldots,v_k,v_1A,\ldots v_kA$ 
and let 
\be
\label{eq:prob1}
\cD(v_1,v_2,\ldots,v_k, v_1A,v_2A,\ldots,v_kA)
\ee
be the joint distribution of the $2k$-tuple, which can be viewed as the pushforward 
$(T_{k,A})_*\mu$ of the measure $\mu$ under the map 
\benn
T_{k,A}(x)=(v_1(x),v_2(x),\ldots,v_k(x), (v_1A)(x),(v_2A)(x),\ldots,(v_kA)(x)),
\eenn
where $x\in\Omega$. The $k$-profile $\mathcal{S}_k(A)$ is the set of all probability 
measures of the form~\eqref{eq:prob1}, where we go through all possible $k$-tuples of 
functions in $L^\I(\Omega,[-1,1])$. In summary, the two crucial ideas are to view 
more analytically graphs as \emph{operators} and to use profiles to compare 
matrices/graphs/P-operators \emph{regardless of their size}. Let 
$A,B$ be two P-operators. One defines 
a metric
\begin{equation}
d_M(A,B):=\sum_{k=1}^\infty 2^{-k} d_H(\mathcal{S}_k(A),\mathcal{S}_k(B)),
\end{equation}
where $d_H$ is the Hausdorff distance for $X,Y\in\mathcal{P}(R^k)$ (i.e., $X,Y$ 
are subsets of probability measures on $\R^k$)
\begin{equation}
d_H(X,Y):=\max\left\{ \sup_{x\in X}\inf_{y\in Y} d_{LP}(x,y), 
\sup_{y\in Y}\inf_{x\in X} d_{LP}(x,y)\right\}
\end{equation}
and $d_{LP}$ is the L\'evy-Prokhorov metric. This metric is actually quite 
natural and given for $\nu_1,\nu_2\in\R^{k}$ by
\begin{equation}
d_{LP}(\nu_1,\nu_2):=\inf\{\varepsilon>0:\nu_1(U)\leq \nu_2(U^\varepsilon)
+\varepsilon\text{ and }\nu_2(U)\leq \nu_1(U^\varepsilon)+\varepsilon\text{ 
$\forall U\in\mathcal{B}_k$}\},
\end{equation}
where $\mathcal{B}_k$ is the Borel sigma-algebra on $\R^k$ and $U^\varepsilon$ 
is the set of points having distance less than $\varepsilon$ from $U$. It is 
well-known that convergence in the L\'evy-Prokhorov metric essentially is equivalent 
to weak convergence of measures so to actually ensure the existence of non-trivial 
limits, it is well-balanced choice between retaining some complexity in the limit,
yet still having some form of mean-field object.\medskip

The definition of P-operators generalizes in a natural way to 
$\|A\|_{p\rightarrow q}$. One may metrize the space of P-operators via
\be
\label{eq:Popmet}
d_M(A,B):=\sum_{k=1}^\infty 2^{-k} d_H(\mathcal{S}_k(A),\mathcal{S}_k(B)),
\ee
and call the associated convergence notion \emph{action convergence}. Then one can 
prove~\cite{BackhauszSzegedy} that any sequence $\{A^(j)\}_{j=1}^\infty$ with uniformly 
bounded norm $\|\cdot\|_{p\rightarrow q}$ for $p\in[1,\I)$ and $q\in[1,\I]$ has a 
limit~\cite[Thm.~2.14]{BackhauszSzegedy}. This convergence is seen to generalize the 
theory of graphons and even graphings (classical objects used for very sparse graphs) 
as well as many intermediate cases. For example, if we have a graphon~\cite{Lovasz} 
$W:\Omega\times \Omega\rightarrow \R$ with finite norm
\be
\label{eq:graphon}
\|W\|_p:=\int_{\Omega\times \Omega}W(x,y)~\textnormal{d} \mu^2, 
\quad p\in[1,\infty),~q:=\frac{p}{p-1},
\ee  
then we can define an associated P-operator by
\be
\label{eq:graphonPop}
(fA_W)(x):=\int_{\Omega} W(y,x)f(y)~\textnormal{d}\mu.
\ee
It is easy to see that the norm $\|A_W\|_{p\rightarrow q}$ is finite in this case 
and $A_W$ is just an operator-theoretic viewpoint on the $L^p$-graphon $W$. Again, 
compactness results hold for the case of $\|\cdot\|_{p\rightarrow q}$ so that we 
get the existence of a limiting graphon/P-operator.\medskip

There are also special classes of P-operators, which are particularly 
relevant. One important class are \emph{graphops}, which are positivity 
preserving and self-adjoint P-operators, which behave effectively like 
undirected graphs. This makes sense since in the finite-dimensional case, adjacency 
matrices of undirected graphs are \emph{graphops}. 

Once we have graphops, it is natural to ask, how these operators are going to
appear in dynamical systems on networks. It is the main theme of this paper to
understand their appearance in dynamics on a formal level (rigorous proofs seem
to be out of reach in full generality at this point but it is expected that 
eventually such proofs will be available).

\section{Microscopic Network Dynamics}
\label{sec:micro}

We always denote the dynamical state of a vertex by $u_k(t)$ for $k\in[n]$ and denote the
current time by $t\in[0,T)$ with some fixed $T>0$.

\subsection{The Kuramoto Model}
\label{ssec:Kura1}

The classical Kuramoto model considers oscillators on a circle so 
$u_k(t)\in \mathbb{S}^1:=\R/(0\sim 2\pi)$. The dynamics is given by
\be
\label{eq:Kuraclassical}
\frac{\txtd u_k}{\txtd t}=:u'_k=\omega_k+\frac{p_0}{n}\sum_{j=0}^n \sin(u_j-u_k),
\ee
where $\omega_k$ are given internal frequencies of the individual oscillators
and $p_0\geq 0$ is a parameter controlling the coupling. The Kuramoto model on
complex networks is usually written as follows~\cite{Rodriguesetal2}
\be
\label{eq:KuraComplex}
u'_k=\omega_k+p\sum_{j=0}^n A_{kj}\sin(u_j-u_k),
\ee
where $p\geq 0$ is a parameter controlling the coupling, which may contain
a certain scaling in $n$ depending upon the type of graph defined by the adjacency
matrix $A=(A_{kj})_{k,j\in[n]}$. Of course, on a finite-dimensional level the  
interpretation of adjacency matrices as graphops on $([n],2^{[n]},\mu_{[n]})$, as 
discussed in Section~\ref{sec:graphops}, applies. 
So if we consider a matrix of phase differences
\benn
v_{jk}:=\sin(u_j-u_k),\qquad V:=(v_{jk})_{j,k\in[n]},
\eenn
this yields upon inserting it into the Kuramoto model
\be
\label{eq:KuraComplex1}
u_k'=\omega_k+p~\textnormal{diag}(AV),
\ee
where $\textnormal{diag}(M)$ is the vector obtained from the elements on the
diagonal of the matrix $M$. Note that $(V,AV)$ contains the same information as
\benn
(v_{\cdot 1}^\top,v_{\cdot 2}^\top,\ldots,v_{\cdot n}^\top,
v_{\cdot 1}^\top A,v_{\cdot 2}^\top A,\ldots,v_{\cdot n}^\top A)
\eenn
where we used $A=A^\top$ as our graphs are undirected. Hence, $(V,AV)$ can be 
viewed as an $n$-profile if we uniformly sample from
it. Since the matrix of phase differences $V$ does \emph{evolve in time},
we see that the input to the dynamics of each oscillator is (in addition 
to its intrinsic frequency) driven by moving through a subset of the 
$n$-profiles. Even beyond identifying the driving, we can go further by setting
\benn
\omega=(\omega_1,\ldots,\omega_n),\quad u=(u_1,\ldots,u_n),\quad
d=((AV)_{11},\ldots,(AV)_{nn}),
\eenn
to re-write our differential equations as $u'=\omega+pd$, so right-multiplication
and time-differentiation of the $1$-profile $(u,uA)$ gives
\benn
(u,(uA))=(u',(uA)')=(\omega+pd,\omega A,+pdA).
\eenn
Therefore, this defines an evolution equation for the pair $(u,uA)$,
which yields an evolution equation on the space of \emph{measures} $(T_{1,A})_*\mu$
also written by
\be
\label{eq:jed}
\cD(u,uA)=\frac1n\sum_{j=1}^n \delta_{u_j,(uA)_j}
\ee
which are associated to a $1$-profile. In summary, the dynamics of the Kuramoto
model on finite graphs induces an evolution equation on measures, which is somewhat
different from using the classical empirical measure
\benn
\frac1n\sum_{j=1}^n \delta_{u_j},
\eenn
which keeps only track of the positions. Indeed, the $1$-profile also contains the 
action of $A$ on $u$ given by $uA$ and thereby the relevant \emph{correlation structure} 
embodied by the \emph{joint empirical 
distribution} (aka the``JEDi'') given by~\eqref{eq:jed}. The importance of this 
novel viewpoint of the dynamics is that once we are on a graph, then we have to 
intertwine structure and dynamics~\cite{KuehnNetworks1}, thereby capturing how 
vertices respond to connectivity. We will see that this theme re-appears on other
modelling scales and for VFPEs below.

\subsection{The Cucker-Smale Model}

The Cucker-Smale model~\cite{CuckerSmale1} is one well-known model for swarming. One 
variant of it reads
\benn
u_k''=\frac1n \sum_{j=1}^n (u_j'-u_k') \psi(|u_k-u_j|),
\eenn
where $u_k=u_k(t)$ is the position of agent $k$ at time $t$ and $\psi$ is a
given function regulating the type of communication between different agents. 
More generally, one may pose the model on a network via 
\benn
u_k''=\frac1n \sum_{j=1}^n a_{kj}(u_j'-u_k') \psi(|u_k-u_j|).
\eenn
We set $\tilde{a}_{kj}:=a_{kj}\psi(|u_k-u_j|)$, and obtain by re-writing the 
model in vectorized form as a first-order system and by right-multiplication with
$A$, the following set of equations
\be
\begin{array}{lcl}
u'&=&v,\\
v'&=& \frac1n\left[\tilde{A}v-\textnormal{Diag}(\tilde{A}\tilde{A}^\top)u\right],\\
(uA)'&=&vA,\\
(vA)'&=& \frac1n\left[\tilde{A}(vA)-\textnormal{Diag}(\tilde{A}\tilde{A}^\top)(uA)\right],
\end{array}
\ee
where $\textnormal{Diag}(M)$ denotes the matrix obtained from $M$ by only keeping the
entries on the diagonal and setting all other entries to zero.
Of course, the equations for $(uA,vA)$ still depend directly on values of $u$ via the
definition of $\tilde{A}$. It is interesting to see that we effectively obtain now
an evolution of measures via the $2$-profile $(u,v,uA,vA)$. The structure of the equations 
entails that the second-order nature of the model transfers to profiles, i.e., 
there is a constraint on the $2$-profile in the same sense as for usual second-order
ODEs.   

\subsection{Kinetic Models}
\label{ssec:kinbase}

Instead of particular models, one can also look at broader classes of interacting
systems frequently employed in kinetic theory~\cite{Cercignani,Golse,Spohn1}. A form 
commonly found in the literature is:
\be
\label{eq:kingen}
u_k'=\sum_{j\neq k} f(u_j,u_k).
\ee
Evidently, one can extend this to a model with a complex network structure by writing
\be
\label{eq:kingen1}
u_k'=\sum_{j=1}^n a_{kj}f(u_j,u_k).
\ee 
In this general abstract form, if we set $f(u_j,u_k)=:f_{jk}=f_{jk}(u)$ we get with a matrix 
$F(u)=(f_{jk}(u))$ that 
\benn
u'=\textnormal{diag}(AF(u))\qquad\text{and}\qquad (uA)'=\textnormal{diag}(AF(u))A,
\eenn
as the evolution equation via the $1$-profile $(u,uA)$. Of course, this formulation is far 
too general to see any special structure regarding the evolution equation induced on the measures 
contained in profiles. If we linearize the evolution equations around a steady state $u_*$ we obtain
\benn
U'=\txtD_u\left[\textnormal{diag}AF(u)\right]_{u=u_*} U \quad\text{and}\quad 
(UA)'=\txtD_u\left[\textnormal{diag}AF(u)\right]_{u=u_*}UA.
\eenn 
So locally near steady states the evolution equation for the $1$-profile are identical for
both components, regardless of the precise underlying kinetic/particle model. This already
hints at the fact, that profiles and the operator-theoretic viewpoint via graphops is 
well-suited for nonlinear dynamics and functional analysis techniques. This viewpoint will
re-appear again for VFPEs below.

\section{Mesoscopic Network Dynamics}
\label{sec:meso}

Having observed that graphops and profiles can provide a very elegant re-interpretation as 
well as augmentation of finite-dimensional network dynamics models, we now proceed to formally
take limits $n\ra \I$ to obtain ``mean-field'' models. We consider the mesoscale, in the
sense that we are no longer interested in the state of individual vertices on a graph but only
in the probability distribution of the state of a vertex. 

\subsection{The Kuramoto Model}
\label{ssec:Kura2}

We have already seen in Section~\ref{ssec:Kura1} that the classical Kuramoto model provides a good
starting point. It is well-studied, and its homogeneous mean-field limiting equation is
well-known. We recall its classical formal derivation~\cite{Kuramoto,Strogatz1} here because this 
derivation will be important for us to reflect back upon in the context of graphops below. 

\subsubsection{All-to-All Coupling}
\label{sec:alltoall}

Starting from~\eqref{eq:Kuraclassical}, suppose we have very large number of oscillators and their
intrinsic frequencies are distributed according to a density $g=g(\omega)$ with 
$g(\omega)=g(-\omega)$. One introduces the complex order parameter as
\be
\label{eq:orderpara}
r\txte^{\txti \psi} := \frac1n \sum_{j=1}^n \txte^{\txti u_j}.
\ee
Multiplying the order parameter by $\txte^{\txti u_k}$, and taking imaginary parts, one 
easily checks that now the Kuramoto model~\eqref{eq:Kuraclassical} can be re-written as
\be
\label{eq:onefeelsall}
u_k'=\omega_k+p_0r\sin(\psi-u_k).
\ee
In particular, the $k$-th oscillator feels all other oscillators via a single mean-field
parameter. Next, let 
\benn
\rho(u,t,\omega)~\txtd u
\eenn
denote the fraction of oscillators with frequency $\omega$ between $u$ and $u+\txtd u$ at time $t$.
Then by construction $\rho$ is a probability density 
\benn
\int_{0}^{2\pi} \rho(u,t,\omega)~\txtd u = 1,\qquad \rho\geq 0.
\eenn
In the limit $n\ra \I$, we can formally re-write the order parameter~\eqref{eq:orderpara} as
\be
\label{eq:opcont}
r(t)\txte^{\txti \psi(t)} = \int_0^{2\pi} \int_\R \txte^{\txti u} 
\rho(u,t,\omega)g(\omega)~\txtd \omega~ \txtd u.
\ee 
Indeed, the last formula can be derived as a limit of the sum in~\eqref{eq:orderpara} using
the law of large numbers~\cite{Strogatz1}. Yet, it is very crucial to note that this law
of large numbers argument only gives us a mean-field as it produces only the mean of the
order parameter completely discarding correlations. Furthermore, the approach implicitly 
assumed that each random variable has finite variance. Yet, if these assumptions hold 
then we know that the resulting Liouville/continuity equation for the probability density 
$\rho$ for the Kuramoto model should be given by
\benn
\partial_t \rho = -\partial_u (\rho v),\qquad v(u,t,\omega):=\omega + p_0 r \sin(\psi-u).
\eenn
This equation can be re-written using the order parameter~\eqref{eq:opcont}, multiplying by a
suitable exponential and taking the imaginary part as 
\be
\label{eq:mfKura1}
\partial_t \rho = -\partial_u \left(\rho\left(\omega + p_0 \int_0^{2\pi} \int_\R \sin(\tilde{u}-u) 
\rho(\tilde{u},t,\tilde{\omega})g(\tilde{\omega})\txtd \tilde{\omega} \txtd \tilde{u}\right) \right).
\ee
The previous argumentation can be made fully rigorous to derive the mean-field Vlasov-type 
equation~\eqref{eq:mfKura1}. Yet, the argument is highly non-trivial and the currently most 
elegant way of proof proceeds via a so-called Dobrushin-type bound~\cite{Dobrushin,Dobrushin1}, 
which is effectively a Gronwall estimate on a metric space of measures, usually carried out in 
a Wasserstein metric~\cite{Golse,KuehnBook1,Neunzert}. We remark that if one would consider the 
Kuramoto model with stochastic perturbations, then one is going to obtain a (nonlinear) 
Fokker-Planck equation~\cite{Frank}, which appears in many classical stochastic coupled oscillator 
models, where the coupling appears through a mean field, such as the Desai-Zwanzig 
model~\cite{DesaiZwanzig}. Hence, one sometimes refers to~\eqref{eq:mfKura1} as a
Vlasov-Fokker-Planck (VFPE) equation. 

\subsubsection{Complex Network Heuristics}
\label{sec:heuristics}

Having dealt with the classical all-to-all coupled VFPE case, it is natural to think about 
extensions to treat the Kuramoto model~\eqref{eq:KuraComplex} on complex networks. Recall 
that we have assumed that the underlying graph is undirected and unweighted so that the
adjacency matrix $A=(A_{ij})_{i,j\in[n]}$ is symmetric and binary $A_{ij}\in\{0,1\}$ for all
$i,j\in[n]$. There exists a known formal approach~\cite{Rodriguesetal2} to try to ``save'' the 
classical mean-field argument from Section~\ref{sec:alltoall}. One defines a local order parameter 
as
\benn
r_k\txte^{\txti \psi_k}:=\sum_{j=1}^n A_{kj}\txte^{\txti u_j}.
\eenn
If the graph is sufficiently well-connected and effectively still behaves like an all-to-all
coupled system, one is then tempted to make the ansatz of a single global field
\benn
r\txte^{\txti \psi}=\frac{1}{\kappa_k} r_k \txte^{\txti \psi_k},
\eenn  
where $\kappa_k$ is the degree of vertex $k$, i.e., the global field is locally proportional
to the local field weighted by the degree. With this assumption one obtains 
that~\eqref{eq:onefeelsall} can now be written as
\benn
u_k'=\omega_k+p_0r\kappa_k\sin(\psi-u_k).
\eenn
In particular, these steps motivate that one might want to use the local degree of a vertex
as the next approximation criterion to derive a mean field~\cite{Ichinomiya}. Hence, one
option is to consider a probability density $\rho(u,t,\omega,\kappa)$ of oscillators having
phase $u$, frequency $\omega$, and degree $\kappa$ at time $t$ with the usual conditions
\benn
\int_{0}^{2\pi} \rho(u,t,\omega,\kappa)~\txtd u = 1,\qquad \rho\geq 0.
\eenn
Evidently one cannot really do much with standard tools as just defining $\rho$ still does
not lead to a mean-field VFPE. If one assumes that the network is \emph{uncorrelated} and
has a degree distribution $d(\kappa)$, then the probability that an edge has its end a 
vertex of phase $u$, degree $\kappa$ and frequency $\omega$ at time $t$ is
\benn
\frac{\kappa d(\kappa)}{\langle \kappa \rangle} g(\omega) \rho(u,t,\omega,\kappa)
\eenn
where $\langle \kappa\rangle$ is the average degree of a vertex in the graph. 
This suggests that it might be helpful to define the order parameter now as
\benn
r\txte^{\txti \psi}:=\int_0^{2\pi}\int_\R\int_0^\I
\frac{\kappa d(\kappa)}{\langle \kappa \rangle} g(\omega) 
\rho(u,t,\omega,\kappa)~\txtd \omega~\txtd \kappa ~\txtd u.
\eenn
Now the same trick as previously, multiplying by an exponential and taking
imaginary parts, yields VFPE equation
\benn
\partial_t \rho = -\partial_u \left(\rho (\omega + p \kappa r\sin(\psi-u))\right).
\eenn
The right-hand side of the VFPE could again be expressed now as
some integral. Indeed, one
may formally write the equation for the evolution of the average phase in the limit
$n\ra \I$ as  
\be
u'=\omega+\frac{p\kappa}{\langle \kappa \rangle}\int_0^{2\pi}\int_\R\int_0^\I
g(\tilde{\omega})d(\tilde{\kappa})\tilde{\kappa}\rho(\tilde{u},t,\tilde{\omega},\tilde{\kappa})
\sin(\tilde{u}-u) ~\txtd \tilde{\kappa} ~ \txtd \tilde{u} ~\txtd \tilde{\omega},
\ee
which arises as a formal replacement of the sum in the Kuramoto 
model~\eqref{eq:KuraComplex}. This suggests a closed VFPE in the form  
\be
\label{eq:mfKura2}
\partial_t \rho = -\partial_u \left(\rho \left(
\omega+\frac{p}{\langle \kappa \rangle}\int_0^{2\pi}\int_\R\int_0^\I
g(\tilde{\omega})d(\tilde{\kappa})\tilde{\kappa}\rho(\tilde{u},t,\tilde{\omega},\tilde{\kappa})
\sin(\tilde{u}-u) ~\txtd \tilde{\kappa} ~ \txtd \tilde{u} ~\txtd \tilde{\omega}\right)\right).
\ee
The last considerations already show an emerging theme that we shall exploit later on:
there is an effective integral \emph{operator acting on the density} $\rho$, which depends
upon the graph structure, in the mean-field VFPE equation on complex networks. In
comparison to the classical all-to-all coupled VPDE~\eqref{eq:mfKura1}, we have 
replaced in~\eqref{eq:mfKura2} 
\benn
\rho(\tilde{u},t,\tilde{\omega}) \qquad \textnormal{by} 
\qquad \int_0^\I \frac{d(\tilde{\kappa})\tilde{\kappa}}{\langle
\kappa\rangle} \rho(\tilde{u},t,\tilde{\omega},\tilde{\kappa})~\txtd \tilde{\kappa}.
\eenn  

\subsubsection{Kuramoto on Graphons}
\label{sec:graphons}

Regarding the previous considerations in Section~\ref{sec:heuristics} on integral operators,
it is now no surprise, how the VFPE equation on graphons should look~\cite{ChibaMedvedev}.
Let $W:\Omega\times \Omega\ra \R$ be a graphon as defined in~\eqref{eq:graphon} and recall
that it defines an associated P-operator $A_W$ via~\eqref{eq:graphonPop}. For graphons we
know that we may identify $\Omega=[0,1]$ with the unit interval. So for $x\in \Omega$ one 
now obtains the VFPE
\be
\label{eq:mfKura3} 
\partial_t \rho = -\partial_u \left(\rho \left(
\omega+p\int_0^{2\pi} \int_\R \int_\Omega
g(\tilde{\omega})W(x,\tilde{x})\rho(\tilde{u},t,\tilde{\omega},\tilde{x})
\sin(\tilde{u}-u) ~\txtd \tilde{x} ~ \txtd \tilde{u} ~\txtd \tilde{\omega}\right)\right).
\ee
for a density $\rho=\rho(u,t,\omega,x)$ of oscillators having phase $u$, frequency $\omega$, 
and graphon position $x$ at time $t$. The position $x\in [0,1]$, where we evaluate a graphon has
the natural interpretation in the graph limit as the position/index of a vertex, while $W(x,y)$
provides a limiting version of the adjacency matrix, i.e., how $x$ is connected to $y$. In
comparison to the classical all-to-all coupled VPDE~\eqref{eq:mfKura1}, we have 
replaced in~\eqref{eq:mfKura3} 
\benn
\rho(\tilde{u},t,\tilde{\omega}) \qquad \textnormal{by} \qquad \int_{\Omega} W(x,\tilde{x}) 
\rho(\tilde{u},t,\tilde{\omega},\tilde{x})~\txtd \tilde{x}.
\eenn  
Now the generalization to far more general classes beyond graphons is becoming 
physically evident.

\subsubsection{Kuramoto on Graphops}
\label{sssec:dyngraphons}

Let $A:L^p(\Omega)\ra L^q(\Omega)$ be a graphop as defined in Section~\ref{sec:graphops}.
Suppose it arises as a limit from a sequence of adjacency matrices 
\benn
A=A^{(\I)}=\lim_{n\ra \I} A^{(n)},
\eenn 
where convergence is understood with respect to the metric $d_M$. From the Kuramoto model 
on complex networks~\eqref{eq:KuraComplex} one now formally obtains the VFPE
\be
\label{eq:mfKura4} 
\partial_t \rho = -\partial_u \left(\rho \left(
\omega+p\int_0^{2\pi} \int_\R g(\tilde{\omega})(A^{(\I)}\rho)(\tilde{u},t,\tilde{\omega},x)
\sin(\tilde{u}-u) ~\txtd \tilde{x} ~ \txtd \tilde{u} ~\txtd \tilde{\omega}\right)\right).
\ee
for a density $\rho=\rho(u,t,\omega,x)$ of oscillators having phase $u$, frequency $\omega$, 
and graphop coordinate $x$ at time $t$. The position $x\in \Omega$ is now more abstract, yet
it still has a meaning in the sense that it should present the position/index of a vertex.
Instead of an integral operator representation, for a P-operator, we have to think of the 
action of a graph, which is actually the primary insight that recently arose in the graph
theory literature as discussed in Section~\ref{sec:graphops}. In~\eqref{eq:mfKura4}, we see,
how this insight can be carried over to easily writing abstract VFPE mean-field type equations,
even if the graph is neither all-to-all, nor uncorrelated, nor homogeneous, nor dense, nor
a special sparse graph. In comparison to the classical all-to-all coupled VPDE~\eqref{eq:mfKura1}, 
we have abstractly replaced in~\eqref{eq:mfKura4} 
\benn
\rho(\tilde{u},t,\tilde{\omega}) \qquad \textnormal{by} \qquad (A^{(\I)}\rho)(\tilde{u},t,
\tilde{\omega},\tilde{x}).
\eenn  
Note that this formulation is much cleaner, more general, and much easier to comprehend in 
comparison to other approaches. 

\subsection{Kinetic Models on Graphops}
\label{ssec:kingra}

The approach in the previous section for the Kuramoto model is now relatively 
easy to \emph{formally} generalize to more abstract kinetic models such as~\eqref{eq:kingen1}. 
One simply takes the normal VFPE for~\eqref{eq:kingen}, which would read 
\be
\partial_t \rho = -\partial_u (\rho V(\rho)),
\ee
where $V$ is completely computable from $f$~\cite{Golse}. Then one lifts the action of the 
finite-dimensional adjacency matrices $A^{(n)}$ onto a limiting graphop $A^{(\I)}$, and 
replaces the density in the vector field driven part $V$ of the VFPE 
\be
\label{eq:limitfinally}
\partial_t \rho = -\partial_u(\rho V(A^{(\I)}\rho)),\qquad \rho=\rho(u,t,x),~x\in\Omega.
\ee  
All the operator-theoretic properties of the finite-dimensional graphs $A^{(n)}$, 
which persist in the limit $A^{(\I)}$, actually can now be used to analyze the VFPE. We know
from~\cite{BackhauszSzegedy} that graphops do carry a lot of information via the graph limit 
for large classes of graphs, so we may strongly expect that an analysis of~\eqref{eq:limitfinally} 
can now proceed along classical lines of dynamical systems theory. For example, steady states 
$\rho_*$ will satisfy
\benn
0=\partial_u(\rho_* V(A^{(\I)}\rho_*)).
\eenn
Linearization is possible via the normal Fredholm derivative in many cases. Suppose that 
we have an evolution equation 
\be
\label{eq:evocool}
\partial_t \rho = -\partial_u(\rho V(A^{(\I)}\rho))\qquad \rho=\rho(t)\in\cX,
\ee
in a suitable Banach space $\cX$ where the P-operator part is such that it defines 
a well-defined evolution in $\cX$.
Then the linearization is just formally given by the chain rule and product rule
\benn
\partial_t \Xi = -\partial_u(\Xi V(A^{(\I)}\rho_*) + \rho_* \txtD_\rho V(A^{(\I)}\Xi) )
=:\cL_\I \Xi 
\eenn
for $\Xi=\Xi(t)\in\cX$. The linear
operator $\cL_\I$ can now be analyzed using classical tools from functional analysis
such as spectral theory. As in Section~\ref{ssec:kinbase} on the finite-dimensional
level, the operator $\cL_\I$ carries in an intertwined way the information about 
\begin{itemize}
 \item the shape of the steady state via $\rho_\I$,
 \item the linearization of the coupling function $f$ via $V'$,
 \item the graph spectral information via $A^{(\I)}$.
\end{itemize}
Hence, the graphop viewpoint provides in a completely natural way the opportunity to
lift spectral information from the finite-dimensional setting to the graph limit in
the context of stability analysis.

\section{Outlook}
\label{sec:outlook}

We have shown that the viewpoint of a more abstract operator-theoretic approach is 
extremely elegant to provide a general theory for network dynamics, when we do not
have a simple coupling structure. This applies to the finite-dimensional microscopic
case as well as the limiting case for VFPEs.

Although we have physically derived, by using analogies to previous approaches in the
mean-field context for homogeneous/all-to-all coupling, a suitable VFPE formulation
on a limiting graphop, several questions remain. From a kinetic theory perspective,
we would like to prove that the limiting VFPE indeed approximates over a finite 
time scale, the underlying finite-dimensional network dynamics for large $n$. We
conjecture that similar proofs\footnote{A rigorous proof is currently work in 
progress.} via a Dobrushin-type estimate can also work in the graphop context, 
when the structure of the graph limit theory is taken into account. From the dynamics
perspective, it would be very interesting to study spectral theory for operators
involving graphops as one element of their definition in more detail as we have
indicated in the last section. Furthermore, from the physical perspective it is
important to gain a better understanding of graph limit procedures in the sense 
of particle interactions. We believe that the finite-dimensional
observations we provided hint at a correlation structure viewpoint being 
particularly important to provide the correct physical interpretation. From the 
viewpoint of direct applications in various sciences, it seems reasonable to assume 
that testing examples and analyzing finite-size effects are going to be of particular 
practical importance.\medskip

\textbf{Acknowledgments:} CK acknowledges partial support of the DFG via the SFB/TR 109 
Discretization in Geometry and Dynamics, partial support by the TiPES (Tipping Points in the 
Earth System) project funded the European Union's Horizon 2020 research and innovation 
programme under grant agreement No 820970 and support by a Lichtenberg Professorship of the 
VolkswagenStiftung. CK also thanks the TUM International Graduate School of Science and 
Engineering IGSSE for support via the project SEND (Synchronization in Evolutionary Network 
Dynamics). CK also appreciated general discussions with Sebastian Throm and Tobias B\"ohle
regarding limits of network dynamics.

{
\small
\bibliographystyle{plain}
\bibliography{../my_refs}
}

\end{document}